\documentclass[aps,pra,twocolumn,10pt,longbibliography]{revtex4-1}
\usepackage{graphicx}
\usepackage{amsmath}
\usepackage{amsfonts}
\usepackage{bm}
\usepackage{color}
\usepackage[T1]{fontenc}

\begin{document}

\title{Deterministic magnetization switching by voltage-control
of magnetic anisotropy and Dzyaloshinskii-Moriya interaction under
in-plane magnetic field}

\author{Hiroshi Imamura$^1$}\email{h-imamura@aist.go.jp}
\author{Takayuki Nozaki$^1$}
\author{Shinji Yuasa$^1$}
\author{Yoshishige Suzuki$^{1,2}$}
\affiliation{
$^1$National Institute of Advanced Industrial Science and
Technology(AIST), Spintronics Research Center, Tsukuba, Ibaraki
305-8568, Japan
\\
$^2$Graduate School of Engineering Science, Osaka University, Toyonaka, Osaka 560-8531, Japan}


\begin{abstract}
  Based on the micromagnetic simulations the magnetization switching in
  a triangle magnetic element by voltage-control of magnetic anisotropy
  and Dzyaloshinskii-Moriya interaction under in-plane magnetic field is
  proposed. The proposed switching scheme is not the toggle
  switching but the deterministic switching where the magnetic state
  is determined by the polarity of the applied voltage pulse.
  The mechanism and conditions for the switching are clarified.
  The results provide a fast and low-power writing method for
  magnetoresistive random access memories.
\end{abstract}

\maketitle

%========================================
% Introduction
%========================================
Voltage control of magnetic anisotropy (MA) in a thin ferromagnetic film has
attracted much attention as a key phenomenon for low power spin
manipulation\cite{Weisheit2007,Maruyama2009,Nozaki2010,Shiota2011,Nozaki2014,Lin2014,Amiri2015,Kanai2016,Grezes2016,Munira2016,Nozaki2016,Shiota2016,Nozaki2017}.
The
first experimental demonstration was reported by Weisheit et al. in 2007
\cite{Weisheit2007}. They showed that the MA
of ordered iron-platinum (FePt) and
iron-palladium (FePd) intermetallic compounds can be reversibly modified
by an applied electric field when immersed in an electrolyte.
Two years later Maruyama et al. observed voltage-induced change of
magnetic anisotropy in an all solid state device with an MgO dielectric
layer\cite{Maruyama2009} including MgO-based magnetic tunnel junctions
\cite{Nozaki2010}, which paved the way for the voltage-controlled
magnetic random access memory (VC-MRAM) with low power
consumption\cite{Shiota2011,Nozaki2014,Lin2014,Amiri2015,Kanai2016,Grezes2016,Munira2016,Nozaki2016,Shiota2016,Nozaki2017}.

The writing procedure of a VC-MRAM is as follows.
The memory cell consists of a perpendicularly magnetized recording
layer and is subjected to an in-plane external magnetic field.
The voltage pulse is applied to the cell to remove the MA and induce the
precession of the magnetic moments around the external magnetic field.
If the voltage is turned off at one-half period of precession the
magnetization switching completes. Since this is the toggle-mode
switching preread is necessary; i.e., one has to read the
information stored in the cell before writing.  It is highly desired to
develop another fast and low-power writing method to avoid the cost of
preread.

Recently a growing interest in the interface Dzyaloshinskii-Moriya
interaction (DMI)
\cite{Dzyaloshinskii1960,Moriya1960,Crepieux1998,Bode2007,Ferriani2008,Muhlbauer2009xo,Neubauer2009,Pappas2009,Thiaville2012,Fert2013,Ryu2013,Emori2013o,Torrejon2014,Nawaoka2015,Han2016}
emerged because of its relevance to non-collinear magnetic structures
such as magnetic walls and magnetic skyrmions\cite{Skyrme1962}.
The value of the DMI constant ranges from 0.05 to 1 mJ/m$^{2}$
depending on the materials. In 2015 Nawaoka et al. found
that the interface DMI in the Au/Fe/MgO artificial multilayer can be
controlled by voltage \cite{Nawaoka2015}. Although the voltage-induced
change in the DMI constant is estimated to be as small as 40 nJ/m$^{2}$
at 1 V the observation showed the possibility of manipulating
magnetization by using voltage-induced changes of MA and DMI.

%========================================
% In this paper
%========================================
In this paper we performed micromagnetic simulations and showed
that voltage-induced changes of magnetic anisotropy and interface DMI
can switch the magnetization of a perpendicularly magnetized right
triangle under in-plane magnetic field. The magnetic state after the voltage pulse is determined by the polarity of
the voltage irrespective of the initial magnetic state.

%========================================
% Fig 1 model
%========================================
\begin{figure}[t]
  \includegraphics [width=0.95\columnwidth] {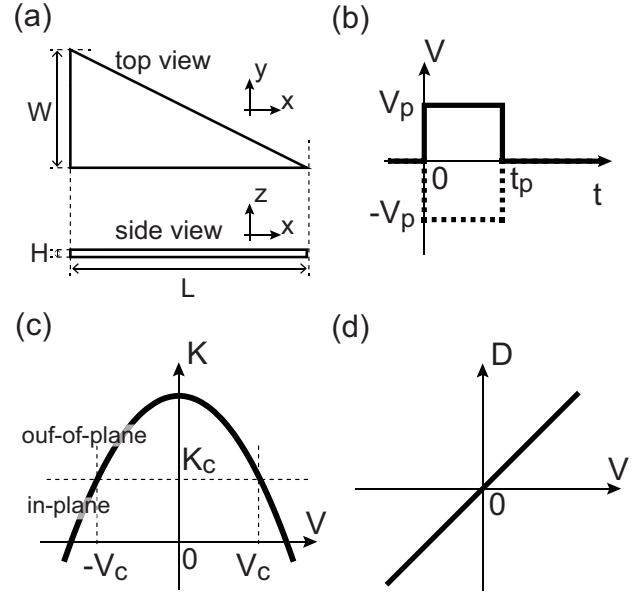}
  \caption{
  \label{fig:model}
  (a) Top and side views of a right triangle element, where the width,
  length and height are presented by $W$, $L$ and $H$, respectively.
  (b) Time ($t$) dependence of the voltage ($V$) for the positive (solid)
  and negative (dotted) voltage pulse with amplitude of
  $V_{\rm p}$,
  where $t_{\rm p}$ denotes the duration of the pulse.
  (c)
  $V$ dependence of the effective
  anisotropy constant $(K)$, where $K_{\rm c}$ denotes the critical value below which the magnetization becomes the almost in-plane polarized state. $\pm V_{\rm c}$ denotes the corresponding critical voltages.
  (d) $V$ dependence of the DMI constant ($D$).
  }
\end{figure}

%========================================
% Basic concept
%========================================
We consider magnetization switching in a perpendicularly
magnetized right triangle nanomagnet shown in Fig. \ref{fig:model}
(a) by application of a voltage pulse shown in Fig. \ref{fig:model}
(b). There are two magnetic states at equilibrium: one is up-polarized
state and the other is down-polarized state.
Both the MA constant ($K$) and the DMI constant ($D$) are assumed to be
voltage-controllable. Voltage dependence of $K$ has been studied by
several groups\cite{Nozaki2010,Amiri2015,Kanai2014,Lin2014}. Some
reported linear voltage dependence \cite{Nozaki2010,Amiri2015} and
others reported non-linear voltage dependence \cite{Kanai2014,Lin2014}.
In this study $K$ is assumed to be a symmetric function of the voltage ($V$)
and decreases with increase or decrease of $V$ as shown in
Fig. \ref{fig:model} (c).
There is a critical value, $K_{\rm c}$, below
which most magnetic moments are aligned in the plane. Since $K$ is assumed be a symmetric function of $V$ the almost
in-plane magnetic state is realized by application of the voltage of
$|V| \ge V_{\rm c}$, where $V_{\rm c}$ is the critical voltage.

It should be noted that the demagnetization field at the corners is much
weaker than that in the middle. Even at $|V| = V_{\rm c}$ the magnetic
moments at the corners have a considerable out-of-plane component.
One can modify the direction of the magnetic moments at the corners
via the voltage-control of $D$. The DMI is the non-collinear
exchange interaction and rotates the magnetic moments. The magnitude of
$D$ represents the period of rotation. The sign of $D$ represents the
direction of rotation; i.e., clockwise or anti-clockwise.
$D$ is assumed to be an anti-symmetric function of $V$ as shown in
Fig. \ref{fig:model} (d). The direction of rotation of
magnetic moments and therefore the direction of magnetic moments at
the corners can be controlled by the sign of $V$.
Although a material having both the symmetric $V$-dependent $K$ and
anti-symmetric $V$-dependent $D$ has not been fabricated yet, it is
important to study the mechanism and conditions for the deterministic
switching based on micromagnetic simulations prior to the experiments.

After turning off the voltage the magnetic moments relax to
the up-polarized or down-polarized state depending on the magnetic
configuration at the corners. In the relaxation process the corners act
as nucleation sites. The switching mode is not the coherent
rotation but the domain wall displacement. To push the domain wall out
of the magnetic element smoothly the magnetic element should have
low symmetry shape like a right triangle shown in Fig. \ref{fig:model}
(a).

%========================================
% Model
%========================================
The micromagnetic simulations were performed by using the software
package MuMax3\cite{Vansteenkiste2014}. We solve the Landau-Lifshitz
equation defined as
\begin{align}
  \label{eq:LLG}
  \frac{\partial \bm{m}}{\partial t}
  =
  -
  \frac{\gamma}{1+\alpha^2}
  \left\{
  \bm{m} \times B_{\rm eff}
  + \alpha
  \left[ \bm{m} \times \left( \bm{m} \times B_{\rm eff} \right)
  \right]
  \right\},
\end{align}
where $\bm{m}$ is the magnetization unit vector, $t$ is time, $\gamma$
is the gyromagnetic ratio and $\alpha$ is the Gilbert damping constant. The effective field
\begin{align}
  \bm{B}_{\rm eff} = \bm{B}_{\rm ext} + \bm{B}_{\rm demag}  + \bm{B}_{\rm exch} +  \bm{B}_{\rm anis}
\end{align}
comprises the external field $\bm{B}_{\rm ext}$, the demagnetizing field
$\bm{B}_{\rm demag}$, the exchange field $\bm{B}_{\rm exch}$ and the
anisotropy field $\bm{B}_{\rm anis}$. The external field is applied in
the positive $x$-direction:
$\bm{B}_{\rm ext} = B_{\rm ext} \bm{e}_{x}$, where $\bm{e}_{i}$, $(i =
x, y, z)$ represents the unit vector in the direction of the positive
$i$-axis.
The definition of the Cartesian coordinates is given in
Fig. \ref{fig:model} (a).
The anisotropy field is given by
\begin{equation}
  \bm{B}_{\rm anis} = \frac{2 K}{M_{\rm s} H} m_{z}\, \bm{e}_{z},
\end{equation}
where $M_{\rm s}$ is the saturation magnetization and $H$ is the height of the triangle element shown in Fig. \ref{fig:model} (a). The exchange field comprises the Heisenberg exchange interaction term and the DMI
term as
\begin{align}
  \bm{B}_{\rm exch}
  =
  &  \frac{2 A}{M_{\rm s}} \nabla^{2} \bm{m} \notag\\
  &+ \frac{2D}{M_{\rm s}}
  \left(
  \frac{\partial m_z}{\partial x},\ \frac{\partial m_z}{\partial y},\ -\frac{\partial m_x}{\partial x}-\frac{\partial m_y}{\partial y}\right),
\end{align}
where $A$ is the exchange stiffness constant for the Heisenberg interaction.

%========================================
% Fig 2 mx, my, mz
%========================================
\begin{figure}[t]
  \includegraphics [width=0.95\columnwidth] {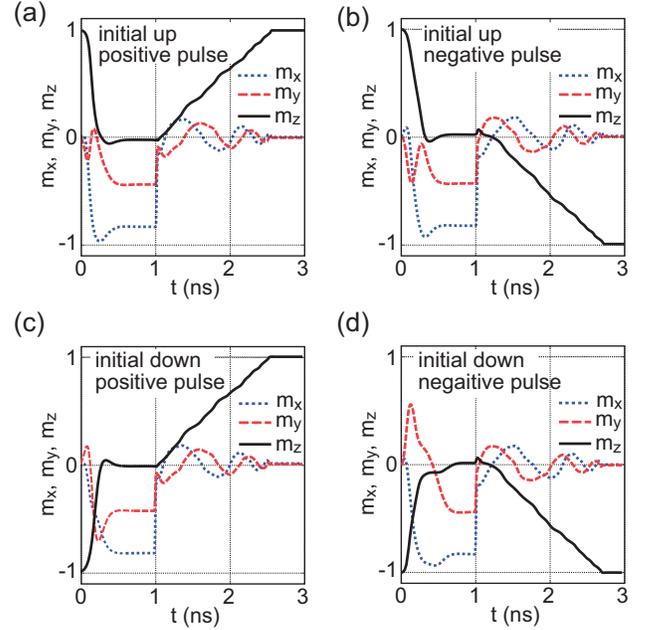}
  \caption{
  \label{fig:mxyz}
  Time evolution of the $x$-, $y$- and $z$-components of the space
  averaged magnetization unit vector, $m_{x}$, $m_{y}$ and $m_{z}$.
  (a) The results for the up-polarized initial state with positive
  voltage pulse ($K_{\rm p}$ = 1.4 mJ/m$^{2}$, $D_{\rm p}$ = 0.5 mJ/m).  The blue dotted, red dashed and black solid
  curves represents $m_{x}$, $m_{y}$ and $m_{z}$, respectively.
  (b) The same plot for up-polarized initial state with negative voltage
  pulse ($K_{\rm p}$ = 1.4 mJ/m$^{2}$, $D_{\rm p}$ = -0.5 mJ/m).
  (c) The same plot for down-polarized initial state with positive
  voltage pulse ($K_{\rm p}$ = 1.4 mJ/m$^{2}$, $D_{\rm p}$ = 0.5 mJ/m).
  (d) The same plot for down-polarized initial state with negative
  voltage pulse ($K_{\rm p}$ = 1.4 mJ/m$^{2}$, $D_{\rm p}$ = -0.5 mJ/m).
  }
\end{figure}

The width, length and height of the triangle element are assumed to be
$W$ = 32 nm, $L$ = 64nm and  $H$ = 2 nm, respectively.
The system is divided into cubic cells of side length 2 nm.
The following material parameters are assumed:
saturation magnetization $M_{\rm s}$ = 1.35 MA/m, exchange stiffness
constant $A$ = 10 pJ/m.
The Gilbert damping constant $\alpha$ ranges from 0.1 to 1.
The MA constant during the pulse ($K_{\rm p}$) ranges from 0 to 2 mJ/m$^{2}$.
The MA constant at $V$ = 0 is assumed to be 4 mJ/m$^{2}$.
The DMI constant during the pulse ($D_{\rm p}$) ranges from 0.01 to 2 mJ/m.
The DMI constant at $V$ = 0 is assumed to be 0.
The shape of the voltage pulse is shown in Fig. \ref{fig:model} (b), where $V_{\rm p}$ and $t_{\rm p}$ are the amplitude and the duration of the pulse.

%========================================
% mx, my, mz
%========================================
Figures \ref{fig:mxyz} (a)-(d) show the typical examples of the
magnetization dynamics. The value of $\alpha$ is assumed to be 1 to
suppress the precessional motion during the pulse.
The $x$-, $y$- and $z$-components of the space
averaged magnetization unit vector are plotted by the blue dotted,
red dashed and black solid curves, respectively.
The external magnetic field of $B_{\rm ext}$=10 mT is applied in the
positive $x$-direction. The initial state is the up-polarized
state for Figs. \ref{fig:mxyz} (a) and (b), and is the down-polarized
state for Figs. \ref{fig:mxyz} (c) and (d), respectively. The anisotropy constant during the pulse is assumed to be $K_{\rm p}$ = 1.4 mJ/m$^{2}$. The DMI
constant during the pulse is assumed to be $D_{\rm p}$=0.5 mJ/m ($D_{\rm p}$ = -0.5
mJ/m ) for the positive (negative) voltage pulse.

In all panels $m_{z}$ is almost zero at the end of the pulse ($t$ = 1
ns) which means that the almost in-plane polarized magnetic state is realized.
After turning off the voltage $m_{z}$ increases or decreases with
increase of $t$ depending on the sign of the voltage.
For the up-polarized initial state ($m_{z}$=1) the magnetization is not
switched by the positive voltage pulse but is switched to the
down-polarized state ($m_{z}$=-1) by the negative voltage pulse as shown
in Figs. \ref{fig:mxyz} (a) and (b).  For the down-polarized initial
state ($m_{z}$=-1) the magnetization is not switched by the negative
voltage pulse but is switched to the up-polarized state ($m_{z}$=1)
by the positive voltage pulse as shown in Figs. \ref{fig:mxyz} (c) and
(d). These results clearly show that the magnetization direction of the
final state is determined by the polarity of the voltage pulse
irrespective of the initial state.

%========================================
% Fig 3 snap shots
%========================================
\begin{figure}[t]
  \includegraphics [width=0.95\columnwidth] {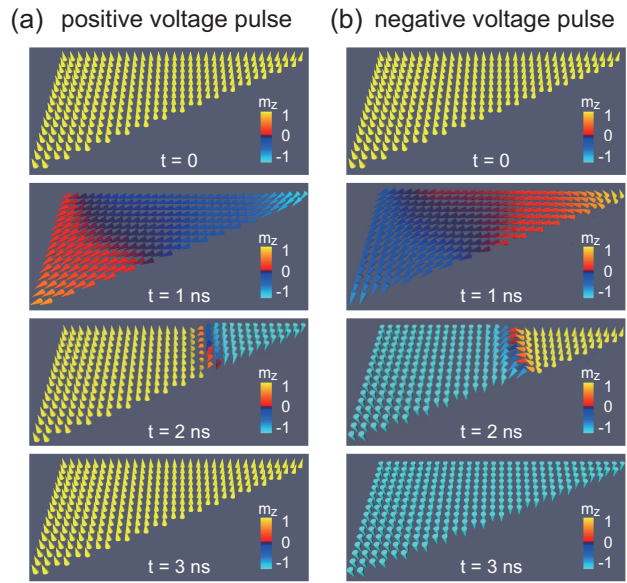}
  \caption{\label{fig:snapshots}
  Snapshots of the magnetization vectors for the up-polarized initial state.
  (a) The snapshots for the positive voltage pulse. From top to bottom
  $t$=0, 1, 2, 3 ns. The red-tone represents the positive $m_{z}$ while
  the blue-tone represents the negative $m_{z}$.
  (b)The same plot for the negative voltage pulse.
  }
\end{figure}

Figures \ref{fig:snapshots} (a) and (b) show the snapshots of the
magnetization vectors at $t$=0, 1, 2, 3 ns for the up-polarized initial
state. The sign of the voltage pulse is positive for
Fig. \ref{fig:snapshots} (a) and negative for Fig. \ref{fig:snapshots}
(b). The red-tone represents the positive $m_{z}$ while the blue-tone
represents the negative $m_{z}$.
The magnetization configuration at the end of the pulse ($t$ = 1 ns) is
quite different between the positive and negative voltage pulses. For
the positive voltage pulse $m_{z}$ at the bottom left corner is positive and
that at the top right corner is negative as shown in the second panel of
Fig. \ref{fig:snapshots} (a). For the negative voltage pulse $m_{z}$ at the
bottom left corner is negative and $m_{z}$ at the top right corner is
positive as shown in the second panel of Fig. \ref{fig:snapshots} (b).

For the positive voltage pulse up-polarized initial state does not
switch to the down-polarized sate but returns to the up-polarized
state. Once the voltage is turned off the magnetic moments around the
bottom left corner tilt upwards to form an up-polarized domain, and those
around the top right corner tilt downwards to form a down-polarized
domain as shown in the third panel of Fig. \ref{fig:snapshots} (a).
Between the up-polarized and down-polarized domains there appears a
narrow domain wall which moves to the top right corner to
reduce the dmain wall energy. Finally the domain wall is swept out from
the magnet, and the up-polarized state is recovered within 3 ns.

Application of the negative voltage pulse switches the magnetization
from up-polarized state to the down-polarized state as shown in
Fig. \ref{fig:snapshots} (b). After turning off the voltage pulse the
magnetic moments around the bottom left corner tilt downwards to form a
down-polarized domain, and those around the top right corner tilt
upwards to form an up-polarized domain as shown in the third panel of
Fig. \ref{fig:snapshots} (b). The domain wall between these two domains
moves to the top right corner, and the magnetic state is switched to the
down-polarized state within 3 ns.

%========================================
% Fig 4 mxyz at the end of pulse
%========================================
\begin{figure}[t]
  \includegraphics [width=0.95\columnwidth] {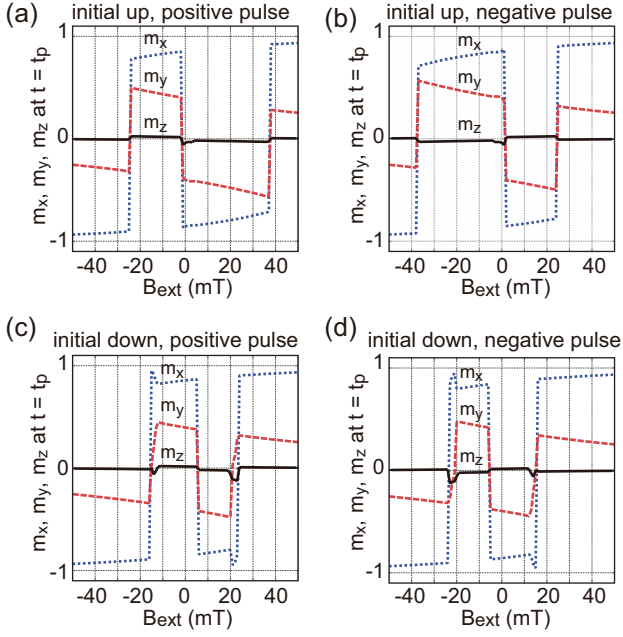}
  \caption{\label{fig:mxyz_at_tp}
  The $x$-, $y$- and $z$-components of the space averaged magnetization unit vector, $m_{x}$, $m_{y}$ and $m_{z}$ at the end of the pulse.
  (a) The results for the up-polarized initial state with positive
  voltage pulse ($K_{\rm p}$ = 1.4 mJ/m$^{2}$, $D_{\rm p}$ = 0.5 mJ/m).  The blue dotted, red dashed and black solid
  curves represents $m_{x}$, $m_{y}$ and $m_{z}$, respectively.
  (b) The same plot for up-polarized initial state with negative voltage
  pulse ($K_{\rm p}$ = 1.4 mJ/m$^{2}$, $D_{\rm p}$ = -0.5 mJ/m).
  (c) The same plot for down-polarized initial state with positive
  voltage pulse ($K_{\rm p}$ = 1.4 mJ/m$^{2}$, $D_{\rm p}$ = 0.5 mJ/m).
  (d) The same plot for down-polarized initial state with negative
  voltage pulse ($K_{\rm p}$ = 1.4 mJ/m$^{2}$, $D_{\rm p}$ = -0.5 mJ/m).
    }
\end{figure}

As shown in the second panels of Figs. \ref{fig:snapshots} (a) and  \ref{fig:snapshots} (b) the most magnetizaion vectors at the end of the pulse point to the negative $x$-direction in spite that $B_{\rm ext}$ of 10 mT is applied in the positive $x$-direction, which implies that 10 mT is not large enough to align the magnetization vectors along $B_{\rm ext}$.
Figures \ref{fig:mxyz_at_tp} (a) --  \ref{fig:mxyz_at_tp} (d) show the $B_{\rm ext}$-dependence
 of $m_{x}$, $m_{y}$ and $m_{z}$ at the end of the pulse.  The parameters other than $B_{\rm ext}$ are the same as those in Figs. \ref{fig:mxyz} and \ref{fig:snapshots}. The magnetization configuration and therefore
 the values of $m_{x}$, $m_{y}$ and $m_{z}$ suddenly change at certain values of $B_{\rm ext}$. At large $B_{\rm ex}$ ($|B_{\rm ex}|\ge$ 40 mT) $m_{x}$ has the same sign as $B_{\rm ext}$; i.e. the most magnetization vectors are aligned long the external magnetic field.  Comparing Fig. \ref{fig:mxyz_at_tp} (a) with \ref{fig:mxyz_at_tp} (b), or \ref{fig:mxyz_at_tp} (c) with \ref{fig:mxyz_at_tp} (d), we note that the magnetization configuration at the end of the pulse has the symmetry with respect to the following transformation: $\{B_{\rm ext}, D_{\rm p}, m_{x}, m_{y}\}$ $\to$
$\{-B_{\rm ext}, -D_{\rm p}, -m_{x}, -m_{y}\}$. Assuming that the magnetic moments are placed on a
two-dimensional plane one can easily show that the LLG equation has the same symmetry.

%========================================
% Fig 5 Switching diagram
%========================================
\begin{figure}[t]
  \includegraphics [width=0.95\columnwidth] {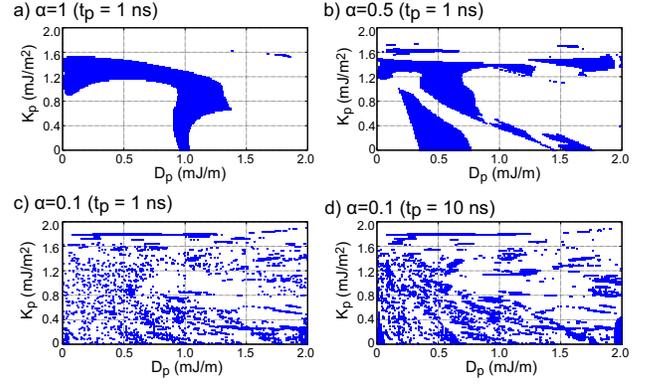}
  \caption{
  \label{fig:swdiagram}
  Switching diagram in terms of  $D_{\rm p}$ and $K_{\rm p}$. The external magnetic field is $B_{\rm ext}$=10 mT.
  (a) The results for $\alpha$ = 1 and $t_{\rm p}$ = 1 ns.
  The regions where the deterministic switching is available are represented by shade.
  (b) The same plot for $\alpha$ = 0.5 and $t_{\rm p}$ = 1 ns.
  (c) The same plot for $\alpha$ = 0.1 and $t_{\rm p}$ = 1 ns.
  (d) The same plot for $\alpha$ = 0.1 and $t_{\rm p}$ = 10 ns.
  }
\end{figure}

Let us discuss the parameter range where deterministic switching is available. Figure \ref{fig:swdiagram} (a) summarizes the simulation results for various values of $D_{\rm p}$ and $K_{\rm p}$ with $\alpha$ = 1. The simulations are performed in the range of 0.01 mJ/m $\le D_{\rm p} \le $ 2 mJ/m and 0 $\le K_{\rm p} \le$ 2 mJ/m$^{2}$. The pulse width is $t_{\rm p}$ = 1 ns and the external magnetic field of $B_{\rm ext}$=10 mT is applied in the positive $x$-direction.
The parameters $(D_{\rm p}, K_{\rm p})$ where the deterministic switching is available are plotted by the blue rectangles most of which bunch around the lines with $K_{\rm p}\simeq 1.4$ mJ/m$^{2}$ and $D_{\rm p}\simeq 1$ mJ/m.
The deterministic switching fails for large $K$ ($\gtrsim 1.6$ mJ/$m^{2}$) because the magnetic moments remains almost perpendicular to the plane at the end of the pulse. At around $K_{\rm p}\simeq 1.4$ mJ/m$^{2}$ the uniaxial anisotropy field is almost canceled by the demagnetization field. Therefore most magnetic moments are aligned in the plane, and the nucleation core at the corners can easily be created by the small value of $D_{\rm p}$.
For small $K$ ($\lesssim 1.2$ mJ/$m^{2}$) the switching region is limited at around $D_{\rm p} \simeq$ 1 mJ/m. As it will be shown later this switching region can be spread down to the lower values of $D_{\rm p}$ by introducing rise and fall time to the square wave pulse.

Figures \ref{fig:swdiagram} (b) and (c) are the same plot for $\alpha$ = 0.5 and 0.1, respectively. For $\alpha=0.5$ the switching region splits into several small pieces and scattered as shown in Fig.  \ref{fig:swdiagram} (b). Also there appear some switching regions at large $K_{\rm p}$ ($\gtrsim 1.6$ mJ/$m^{2}$). Further decrease of $\alpha$ down to 0.1 scatters the switching region into very fine pieces as shown in Fig.  \ref{fig:swdiagram} (c). In Fig. \ref{fig:swdiagram} (d) we plot the results for $\alpha=0.1$ and
$t_{\rm p}$ =10 ns which is long enough for magnetization to relax.
Comparing Fig. \ref{fig:swdiagram} (d) with Fig. \ref{fig:swdiagram} (c) some switching regions are clustered but the difference between them is not significant.  These results imply that the precessional motion of magnetic moments plays an important role in switching failure; i.e. the magnetic moments do not relax into the equilibrium state but into one of the quasi-equilibrium states.

\begin{figure}[t]
  \includegraphics [width=0.95\columnwidth] {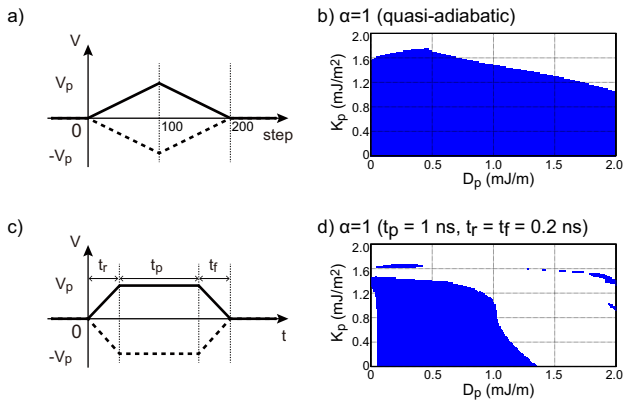}
  \caption{
  \label{fig:rise_fall}
  (a) The quasi-adiabatic evolution of the positive (solid) and negative (dotted) voltage is plotted as a function of the step.
  The absolute value of the voltage is discretized with 100 points; i.e. we take 100 steps to turn on the voltage and another 100 steps to turn off the voltage. At each voltage the micromagnetic simulation is performed until the magnetic moments are relaxed.
  (b) The quasi-adiabatic switching diagram for $\alpha$ = 1.
  (a) Time ($t$) dependence of the voltage $V$ for the positive (solid)
    and negative (dotted) voltage pulse with amplitude of $V_{\rm p}$,
    where $t_{\rm p}$, $t_{\rm r}$ and $t_{\rm f}$ denote the rise time, the duration and the fall time of the pulse, respectively.
  (b) The pulse switching diagram for $\alpha$ = 1, $t_{\rm p}$ = 1 ns, $t_{\rm r}$ = $t_{\rm f}$ = 0.2 ns.
  }
\end{figure}

In order to clarify the effect of the precessional motion on switching we calculate the quasi-adiabatic dynamics of magnetic moments. As shown in Fig. \ref{fig:rise_fall} (a) the absolute value of the voltage pulse $V_{\rm p}$ is discretized with 100 points; i.e.  we take 100 steps to turn on the voltage and another 100 steps to turn off the voltage. At each voltage the micromagnetic simulation is performed until the magnetic moments are relaxed.
Figure \ref{fig:rise_fall} (b) is the switching diagram based on the quasi-adiabatic simulations with $\alpha=1$. Comparing Fig. \ref{fig:rise_fall} (b) with Fig.  \ref{fig:swdiagram} (a) one can clearly see that the failure of switching for $K_{\rm p} \lesssim 1.2$ mJ/m$^{2}$ is due to the precessional motion.

Because we assumed the square shape for the voltage pulse the effective field suddenly changes, and the large precessional torque is exerted on the magnetic moments, at the beginning and end of the pulse. The results shown in Figs. \ref{fig:rise_fall} (a) and (b) suggest that introduction of rise and fall time to the pulse will spread the switching region in Fig. \ref{fig:swdiagram} (a).

%========================================
% Fig 6 Rise and Fall
%========================================

We performed the simulations under the voltage pulse with rise and fall time shown in Fig. \ref{fig:rise_fall} (c). The duration of the pulse is assumed to be $t_{\rm p}$ = 1 ns, and the rise time and the fall time are assumed to be $t_{\rm r}$ = $t_{\rm f}$ = 0.2 ns.  The other parameters are the same as in Fig. \ref{fig:swdiagram} (a).
The switching diagram is shown in Fig. \ref{fig:rise_fall} (d). Introduction of the rise and fall time of 0.2 ns spreads the switching region; i.e. most of the unswitching region in the bottom left part of Fig. \ref{fig:swdiagram} (a) becomes the switching region in Fig. \ref{fig:rise_fall} (d).

%========================================
% Table 2 (B-dep, alpha = 1)
%========================================
\begin{table}[b]
  \begin{tabular}{|c||c|c|c|c|c|c|c|c|c|c|c|}
    \hline $D_{\rm p} \backslash B_{\rm ext}$ &
    0 & 1 & 2 &  3 & 4 & 5 & 6 & 7 & 8 & 9 & 10
    \\
    \hline
    0.01 &
    $\times$ &
    $\times$ &
    $\times$ &
    $\checkmark$ &
    $\checkmark$ &
    $\checkmark$ &
    $\checkmark$ &
    $\checkmark$ &
    $\checkmark$ &
    $\checkmark$ &
    $\checkmark$
    \\
    \hline
    0.02 &
    $\times$ &
    $\times$ &
    $\times$ &
    $\checkmark$ &
    $\checkmark$ &
    $\checkmark$ &
    $\checkmark$ &
    $\checkmark$ &
    $\checkmark$ &
    $\checkmark$ &
    $\checkmark$
    \\
    \hline
    0.05 &
    $\times$ &
    $\times$ &
    $\checkmark$ &
    $\checkmark$ &
    $\checkmark$ &
    $\checkmark$ &
    $\checkmark$ &
    $\checkmark$ &
    $\checkmark$ &
    $\checkmark$ &
    $\checkmark$
    \\
    \hline
    0.1 &
    $\times$ &
    $\times$ &
    $\checkmark$ &
    $\checkmark$ &
    $\checkmark$ &
    $\checkmark$ &
    $\checkmark$ &
    $\checkmark$ &
    $\checkmark$ &
    $\checkmark$ &
    $\checkmark$
    \\
    \hline
    0.2 &
    $\times$ &
    $\times$ &
    $\times$ &
    $\checkmark$ &
    $\checkmark$ &
    $\checkmark$ &
    $\checkmark$ &
    $\checkmark$ &
    $\checkmark$ &
    $\checkmark$ &
    $\checkmark$
    \\
    \hline
    0.5 &
    $\times$ &
    $\times$ &
    $\times$ &
    $\times$ &
    $\times$ &
    $\checkmark$ &
    $\checkmark$ &
    $\checkmark$ &
    $\checkmark$ &
    $\checkmark$ &
    $\times$
    \\
    \hline
  \end{tabular}
  \caption{ \label{table:db}
  $B_{\rm ext}$ and $D_{\rm p}$ dependence of the switching results for $\alpha$ = 1, $K_{\rm p}$=1.4 mJ/m$^{2}$, $t_{\rm r} = t_{\rm f} =0.2$ ns and $t_{\rm p}$ = 1 ns.
  The value of $B_{\rm ext}$ is shown in the unit of mT.  The value of $D_{\rm p}$ is shown in the unit of mJ/m. The check mark indicates that the   deterministic switching successes. The cross indicates that the deterministic switching fails. }
\end{table}

Table \ref{table:db} summarizes the results for various values of
$B_{\rm ext}$ and $D_{\rm p}$. The anisotropy constant during the pulse is assumed to be $K_{\rm p}$=1.4 mJ/m$^{2}$. The other parameters are the same as in Fig. \ref{fig:rise_fall} (d).
Note that the deterministic switching fails for small values of $B_{\rm ext}$.
At $B_{\rm ext}=0$ the LLG equation is invariant under the transformation of $\{D_{\rm p}, m_{x}, m_{y}\}$ $\to$ $\{-D_{\rm p}, -m_{x}, -m_{y}\}$. Because of this symmetry the dynamics of the $z$-component of the
magnetic moments and therefore the magnetic state after the voltage pulse
are the same for both the positive and the negative voltage pulses.
Application of the external field is necessary to break the symmetry
of the LLG equation and make the dynamics of $m_{z}$ different between
the positive and the negative voltage pulses.

%=================
% Summary
%=================
In summary, based on the micromagnetic simulations it is demonstrated
that the voltage-induced changes of MA and DMI can
switch the magnetization of a right triangle magnet under in-plane magnetic
field. The magnetic state after application of the voltage pulse is
determined by the polarity of the voltage irrespective of
the initial magnetic state. The mechanism of switching and the conditions for
the shape of the magnetic element and material parameters are
clarified. The proposed deterministic switching provides the fast and
low-power writing method for VC-MRAMs without preread.

%=================
% Acknowledgement
%=================
This work was partially supported by ImPACT Program of the Council for Science,
Technology and Innovation (Cabinet Office, Government of Japan).

%% \bibliography{refs}

\end{document}